\documentclass[useAMS,usenatbib]{mn2e}
\usepackage{graphicx}
\usepackage{graphicx, amsmath, amssymb}
\title{Gamma Ray Bursts with Extended Emission Observed with BATSE}

\author[Z.F. Bostanc\i, Y. Kaneko, E. G\"o\u{g}\"u\c{s}]
{Zahide Funda Bostanc\i, Yuki Kaneko, Ersin G\"o\u{g}\"u\c{s}$$ \\
  Sabanc\i~University, Faculty of Engineering and Natural
  Sciences, Orhanl\i ~Tuzla 34956 Istanbul Turkey}

\begin{document}
  \date{}
  \pagerange{\pageref{firstpage}--\pageref{lastpage}}
  \pubyear{2012}
  \maketitle
  \label{firstpage}

  \begin{abstract}
We present the results of our systematic search for extended
emission components following initial short gamma-ray burst (GRB)
spikes, using Burst and Transient Source Experiment (BATSE)
observations. We performed the extended emission search for both
short- and long-duration GRBs to unveil the BATSE population of
new hybrid class of GRBs similar to GRB\,060614. For the
identified bursts, we investigate temporal and spectral
characteristics of their initial spikes as well as their extended
emission. Our results reveal that the fraction of GRBs with extended
emission is  $\sim$7\% of the total number of our BATSE sample.
We find that the spectrum of the extended
emission is, in general, softer than that of the initial spike,
which is in accord with what has been observed in the prototypical
bursts, GRB\,060614. We also find that the energy fluence of the
extended emission varies on a broad range from 0.1 to 40 times of
the fluence of the initial spike. We discuss our results in the
context of existing physical models, in particular within the
two-component jet model.
  \end{abstract}


\section{Introduction}
Gamma ray bursts (GRBs) are traditionally divided into long and
short classes, based on the bimodality in the duration distribution
and spectral hardness \citep {kouveliotou93} as observed with the
Burst and Transient Source Experiment (BATSE). However, there is no
clear dividing line for the two classes as their distributions
significantly overlap.  Another ambiguity arises from the fact that
the observed GRB durations could be dilated due to cosmological
redshifts \citep[$\langle z \rangle \sim$ 2$-$3, the most distant so
far with $z \sim 9.4$; e.g.,][]{jakcobsson12, cucc11}, which vary
from burst to burst. Long GRBs, with a duration greater than a few
seconds, are generally distinguished by softer spectra and show
spectral lag, which is the time difference of arrival photon in
separate energy range, while short GRBs are characterized by harder
spectra and have negligible spectral lags \citep{norris00}.

It is mostly believed that the physical origins of long and short
bursts are different although the generation mechanism of GRBs is
still uncertain. Afterglow observations show that long GRBs usually
originate from star forming region in late type galaxies
\citep{fruchter06, levesque10}, and some have accompanying supernovae
likely caused by the core collapse of massive stars
\citep[collapsar;][]{woosley93}. Short GRBs, on the other hand, could
be the results of mergers of binary compact objects, such as two
neutron stars or a neutron star and a black hole \citep{paczynski86,
eichler89, narayan92}. Both early and late types of
host galaxies have been identified for short GRBs and no supernova
association with short GRBs has been observed although they seem to
occur relatively nearby \citep[see][for a review]{berger11}.

The classification of long and short GRBs is further complicated by
the fact that some GRBs exhibit softer, low intensity Extended
Emission (EE) components following the initial short-hard spikes.
Duration of the first component is usually shorter than 5~s, whereas
the second component occurs in time intervals from several seconds up
to $\sim$100~s, usually with gaps of $5-10$~s between the spike and
the EE component. The durations, $T_{90}$\footnote{$T_{90}$ is the
  time during which 90\% of event photons were collected.}, and
subsequent classification of these GRBs then depend on the brightness
and the hardness of the EE components. Such EE has been observed with
various experiments
\citep[e.g.,][]{berth05,villa05,gehrels06}. Specifically, in the Konus
catalog, the EE was found in 11 of short events \citep{mazets02,
  frederiks04}; in the BATSE samples, it was visually identified in 8
bursts, all of which are traditionally classified as long GRBs
\citep{norris06}; and in the second {\it Swift} Burst Alert Telescope
(BAT) catalog, 10 short GRBs with EE were identified
\citep{sakamoto11, norris10}. In addition, long duration burst
  tails (or early afterglows) have been detected in the summed light
  curves of short GRBs in the BATSE data \citep{laz01,
    connaughton02}, in Konus \citep{frederiks04}, and in BeppoSAX
  \citep{montanari05}. Note that these might refer to a different
  phenomenological aspect of short events, although such burst tails
  might result from the superposition of GRBs with weak EE. Since all
these GRBs with EE seem morphologically and spectrally similar, it is
possible that they have the same physical origin regardless of their
classifications based solely on their $T_{90}$ measurements. If
  so, they may comprise a new class of GRBs that possess some
  properties of both short-hard and long-soft GRBs.

A defining example of such a new class GRB is GRB~060614
\citep{gehrels06}. The burst was observed with {\it Swift} BAT
and was a long event ($T_{90} = 102$\,s) as determined with the BAT
in the energy range of 15$-$350\,keV. It contained a short hard
spike with a duration less than 5~s and a soft EE with a duration of
$\sim$100\,s, which was 5 times more energetic than the short spike.
They found that, albeit the long duration, its temporal lag and peak
luminosity fall within the short GRB subclass.  In addition, no
supernova association was detected from this event down to a very
low limit (R-band magnitude $\gtrsim 23$), although it was
originated in rather nearby star-forming galaxy at z = 0.125
\citep{fynbo06, galyam06, dv06}. The original classification of the
long and short GRBs was based on the BATSE GRB sample, which was
sensitive to slightly higher energy ($\gtrsim 30$\,keV) than BAT. In
fact, based on the spectra, BATSE would have identified GRB~060614
with much shorter $T_{90}$ due to the very soft nature of the EE
component. Therefore, the combination of long and short GRB
properties of this burst calls for the new class of GRBs,
which may require a different progenitor scenario than the currently
accepted physical models for long and short GRBs.

Such extended emission could be due to the prolonged activity of the
central engine; however, the long timescale is difficult to reconcile
with the merger model, while the non-association of a supernova in
this case is not in agreement with the standard collapsar model.  At
least for the case of GRB~060614, the EE being an early X-ray
  afterglow is unlikely due to its similarity in spectral lag to the
  initial spike, as well as its high variability \citep{gehrels06}.

In this study, we performed a systematic search for GRBs that are similar to
GRB~060614 using BATSE GRB data archive, to identify GRBs with EE
that may constitute the possible new class. On board the
{\it Compton Gamma Ray Observatory} ({\it CGRO}), BATSE recorded
2704 GRBs, of which $\sim$25\% are classified as ``short" by the
traditional definition of $T_{90} < 2$\,s.  In our search, we
included all GRBs, both short and long. We present the results
of our systematic search, and temporal and spectral properties of
the identified bursts. We organize the paper as follows. In $\S$~2
after a brief description of the BATSE instrument and data types, we
explain the background determination and the search methodology of
our study. In $\S$~3 we give the search results and examine spectral
properties of our findings. Finally, in $\S$~4 we discuss the
results in the context of related physical models.

\section{Data Types and Search Method}

We first give a brief description of the BATSE instrument and of the
data types that we used for our analysis \citep[for a detailed
description, see][]{fis89, pre00, kaneko06}.

\subsection{Data Types}
BATSE consisted of eight identical detector modules located at the
corners of the {\it CGRO} spacecraft for the whole-sky coverage.
Each module had two NaI(Tl) scintillation detectors: a Large Area
Detector (LAD; constant energy coverage of $\sim$30$-$2000 keV) and
a Spectroscopy Detector (SD; variable energy coverage between
$\sim$5 keV and 20 MeV). The data were collected continuously in
low-resolution non-burst mode, and high-resolution burst-mode data
were accumulated when a burst was triggered. They were then
processed in the data processing unit to construct various data
types with different time and energy resolutions.

Although the SD provided a broader energy coverage, the LAD was more
sensitive having a photon collecting area that was 16 times larger
than the SDs. For our search, the data with higher sensitivity are
desirable to identify possibly weak EE components. Therefore, we
used the LAD data of the following types for this study:
Discriminator data (DICSLA), Continuous data (CONT), Time-Tagged
Event data (TTE), Discriminator Science data (DISCSC), and Medium
Energy Resolution data (MER). DISCLA and CONT were continuous,
non-burst mode data. The DISCLA data provided four energy channels
(CH\,1$-$4 corresponding to $\sim$25$-$50, 50$-$100, 100$-$300 and
$>$300\,keV) with 1.024-s time resolution, while CONT data provided
16 energy channels with 2.048-s resolution. TTE, DISCSC and MER were
burst mode data. The TTE and DISCSC data provided four energy
channels (same as DISCLA) and TTE data were time tagged with a
minimum of 2~ms resolution, while DISCSC data had 64-ms time
resolution.  Finally, the MER data provided 16 energy channels with
16~ms time resolution for the first 32.768 s and 64 ms up to
163.84~s after the trigger. We note that the DISCSC and MER data
consist of summed data of 2$-$4 brightest detectors (i.e., detectors
which recorded highest counts from the source), whereas all of the
other data types are for individual detectors.

\subsection{Background Determination and Search Criteria}
For the systematic search for GRBs with EE, we used DISCLA data
collected with the two brightest detectors for each GRB.  We only
considered lowest two channels of DISCLA (25$-$50 and 50$-$100 keV)
since the EE component is usually softer than the initial spike
component.

Determination of background rates plays a critical role in this
study because correct identification of EE component can only be
achieved with the subtraction of accurate background from the burst
light curves. Since the orbit of {\it CGRO} was at the same
geomagnetic coordinates every 15 orbital periods ($\sim$24 hours),
the background rates at a burst trigger time $T_{0}$ could be
approximated by averaging the rates at times $T_0 \pm$15 orbits
\citep{connaughton02}. Therefore, we obtained a background measurement
by using the average of the rates of the two orbits.

\begin{figure}
\centering
\includegraphics[scale=0.4]{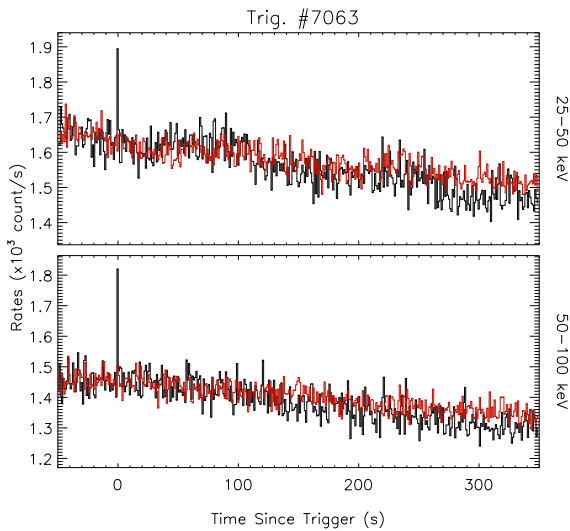}
\caption{Light curves ({\it black line}) of GRB\,980904 (BATSE
Trigger 7063) in two energy bands and orbital backgrounds ({\it red
line}) determined using $T_0\pm15$ orbits.} \label{background}
\end{figure}

We first took into account short GRBs with duration $T_{90}\le$~5~s.
We found 648 such events from the duration table of the Current
BATSE GRB
Catalog\footnote{http://gammaray.nsstc.nasa.gov/batse/grb/catalog/current/},
which contains 2041 bursts. We then subtracted the orbital
background from the observed count rates in each of the two energy
channels. In the case where we could not find the background rates
at times $T_0\pm$15 orbits, we used the rates from $T_0\pm$30 orbits
(2 days before and after), or $T_0\pm$45 orbits.
Figure~\ref{background} shows an example of the orbital background.
There were cases, however, in which no acceptable background was
found either due to data gaps or obvious background mismatches with
the data of $T_0\pm$15, 30, and 45 orbits. These bursts were thus
excluded from our investigation.

\subsection{Search Criteria}
Based on earlier reports of such EE, we focused in the time interval
between $T_0$+5 and $T_0$+350\,s in search of EE.  We binned the
data of this time interval to 4-s resolution and calculated the
Signal to Noise Ratio (SNR) of each bin. We positively identified EE
when SNR $\ge 1.5\sigma$ above the background for at least
consecutive 12 s in both detectors. Inclusion of the second
brightest detector ensures that the EE component detection is not
serendipitous.

Out of 648 short-duration GRBs ($T_{90} \le $ 5 s) in the BATSE
catalog, valid orbital background lightcurves were found only for 269
GRBs. The orbital background data for the other 313 GRBs were
  unavailable, likely due to frequent spacecraft re-orientation.  So
we scanned these events with the above criteria for the EE
search. There are, however, some long GRBs ($T_{90} >$ 5 s) that also
can be classified as a burst with a short spike and EE (e.g., Norris
et al. 2006).  This can occur when the EE components are bright enough
to contribute significantly to the total photon fluence of the
event. To also identify such events in our search, we defined
additional morphological criteria for the 1393 long GRBs ($T_{90} >$ 5
s) in the BATSE catalog: the burst peak should occur before $T_0$+5 s,
and the count rates should remain below 10\% of the peak count rate
for at least 60\% of the duration after the peak time until $T_0$+5
s. These criteria were applied using 64-ms DISCSC light curves (in
25$-$300 keV). The DISCSC data were available for 1373 events out of
the 1393 long GRBs, of which 36 matched the morphological criteria. We
then studied their background data and found that only 18 of them had
valid background data. Consequently, a total of 18 long-duration GRBs
were subjected to the EE search.

Finally, we also considered 663 GRBs without the duration
information available in the catalog. We applied the criteria for
the long GRBs explained above for these events. The DISCSC data were
available for 535 of these GRBs, and of these, we identified 84 GRBs
that matched the morphological criteria; however, due to incomplete
DISCLA data or invalid background, only 9 of them were ultimately
subjected to the further EE investigation.

\section{Search Results and Analysis}

After the systematic search for EE components using 296 BATSE GRBs
with complete data and valid background, we identified a total of 47
GRBs with potential EE components.  We then visually inspected the
energy-resolved light curves of each of these events. We found that
some of these were false detections due to occultation-like steps in
the background (i.e, another bright source being occulted by the
Earth) or inaccurate estimation of the background, and therefore were
discarded from our candidate sample. We also checked their individual
burst reports to exclude GRBs with some known X-ray sources (e.g,
Cygnus X-1) or particle events in the background. As a result, there
were 24 remaining GRBs that consist of short spikes followed by
EE. However, one of them (trig 5989) had EE with three defined peaks,
which resembles a multi-episodic long GRB.  Therefore, we removed it
from our sample. We note that only two of the eight events previously
identified by \citet{norris06} were identified in our search.  This is
because the other six events did not meet our morphological criteria
for long GRBs, or if they did, no valid orbital background data were
available.

\begin{figure*}
  \centering
\includegraphics[scale=0.35]{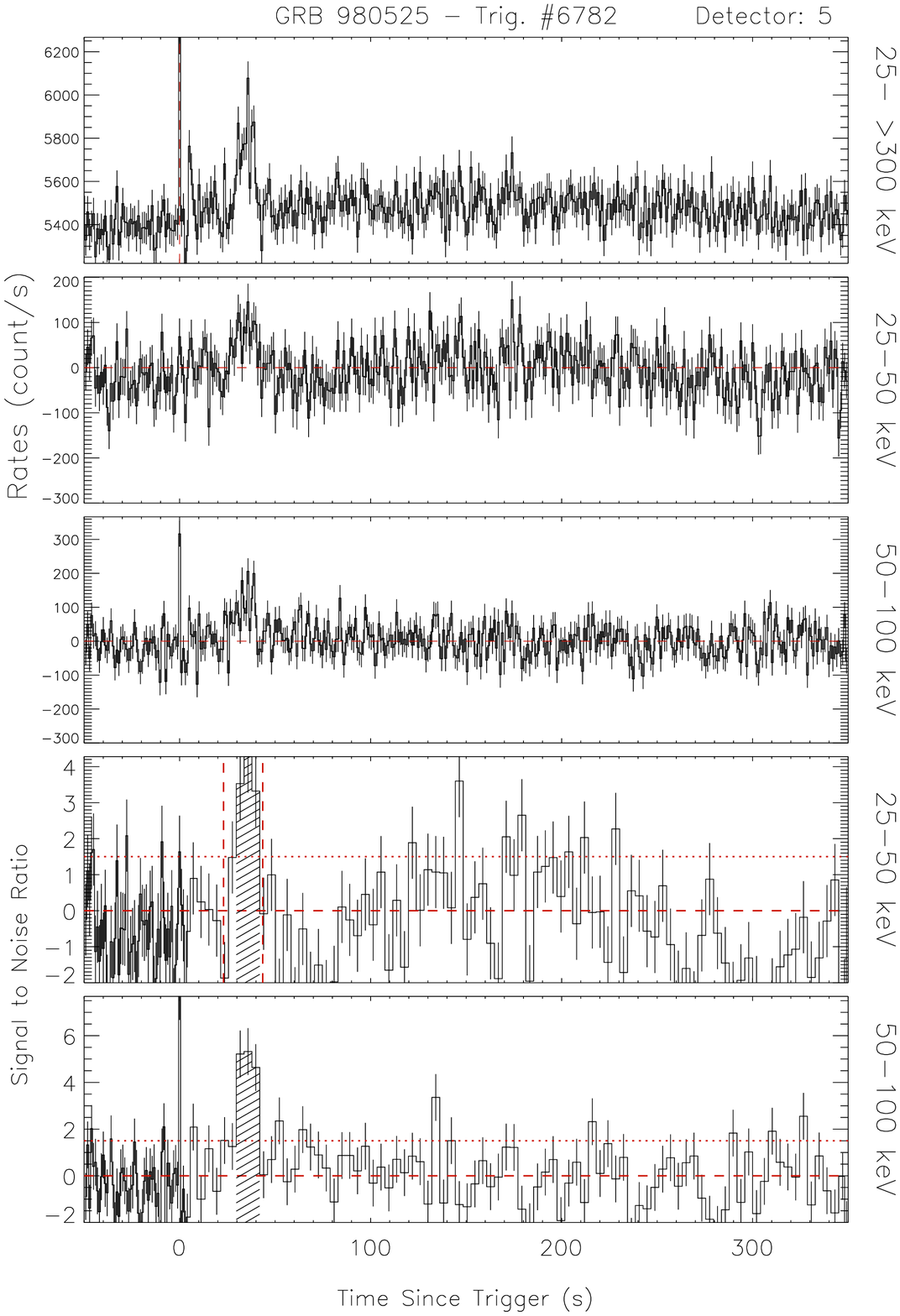}
\includegraphics[scale=0.35]{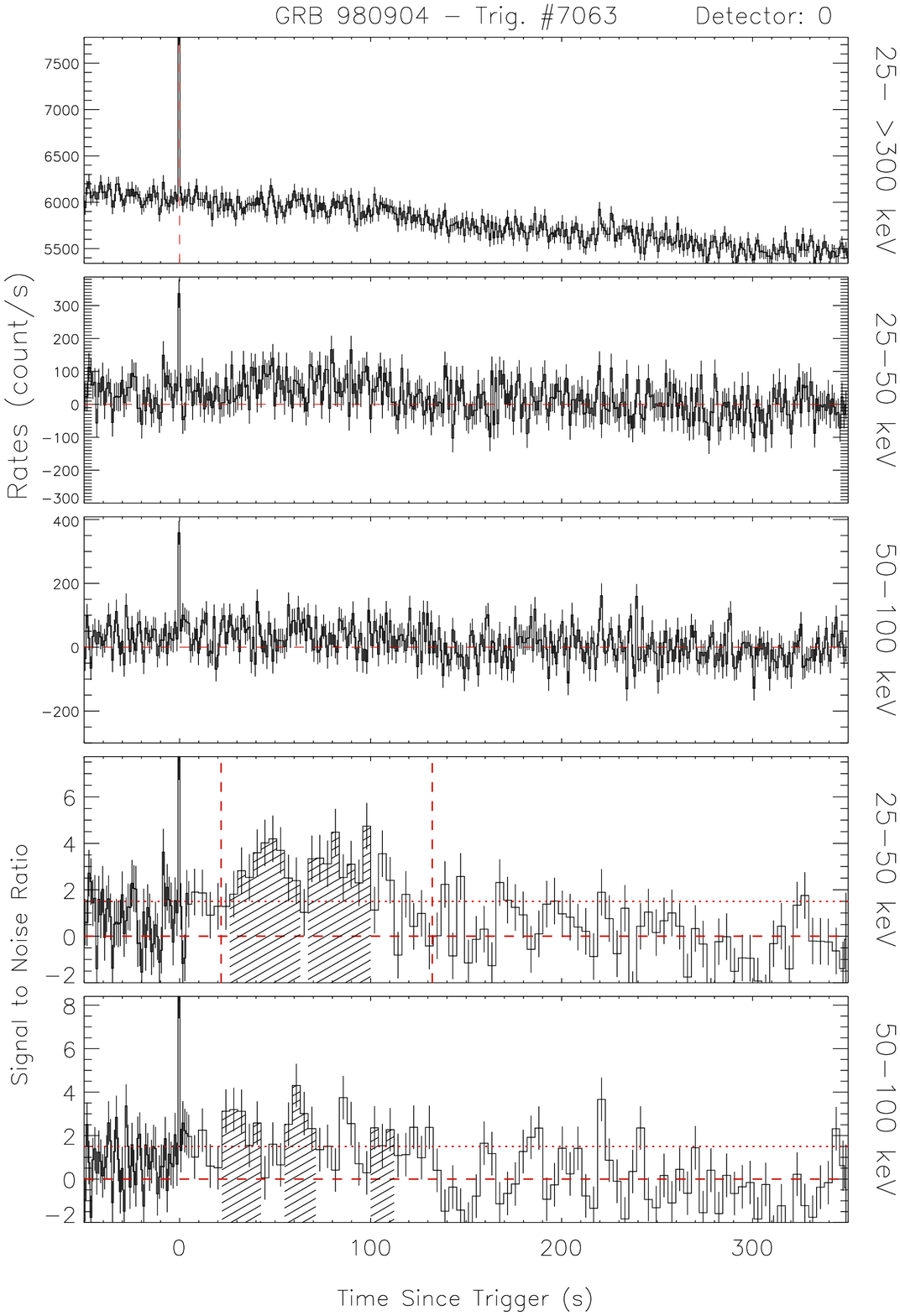}
\caption{Example light curves of the brightest detectors of two GRBs
with EE. For both GRBs, top panel shows sum of four energy channels,
the second and third panels are background subtracted light curves
in two lowest energy channels. The horizontal red dashed line
indicates the level of the background. The last two panels show
signal to noise ratio as a function of time. The shaded areas
indicate the time ranges where the SNR is $\geq$ 1.5$\sigma$ for at
least consecutive 12\,s, and the horizontal dashed and dotted lines
show the background level and the 1.5$\sigma$ level, respectively.
The vertical red dashed lines indicate time interval of EE
component, which was subjected to the spectral analysis.}
\label{s_lightcurves1}
\end{figure*}

Some of the identified EE were very dim, barely above the 1.5$\sigma$
cutoff. In order to perform statistically significant analysis on the
identified EE components, we selected all events with a total SNR
$\geq$ 20 for the entire EE component in CH\,1.  It should be noted
that there were a few cases in which the EE was detected only in
CH\,2, so the SNR criterion was applied to CH\,2 in such cases.

After applying these criteria, we found 19 GRBs with statistically
significant EE. In Figure~\ref{s_lightcurves1}, we present the example
light curves of two GRBs with identified EE components\footnote{All 19
  lightcurves are currently available at
  http://people.sabanciuniv.edu/fbostanci/GRB\_EE/}. Among these GRBs,
11 are short bursts by the traditional definition of $T_{90} < 2$\,s
and seven of them are long events ($T_{90} > 2$\,s).  The durations of
the spikes and the EE range in 0.18--4.86\,s and 12--133 \,s,
respectively (see Table~\ref{duration}).  One of them (GRB 990712,
trig 7647) does not have duration information. We estimated the
duration of this GRB as $T_{90} = 67.8 \pm 2.1$\,s using cumulative
photon fluence, indicating that it belongs to the long. In case more
than one segment of the excess emission were found, we determined the
duration of the detected EE components by assuming that the EE was
continuous from the first segment to the last, based on the lowest
energy light curve (25$-$50\,keV).

\subsection{Spectral Lag Analysis}
In our exploration of the general properties of these GRBs, we first
investigated the spectral lags of their initial spikes. We used the
four-channel TTE data binned to 8-ms resolution in the time range
from $T_0-0.2$ to $T_0+5$~s, and computed the spectral lag of the
signals in CH\,1 with CH\,3 by using a cross correlation function
\citep[see e.g.,][]{norris00}. For GRBs 970918 and 980525 the
spikes were not significant enough in CH\,1; therefore, for these
bursts we used CH\,2 and CH\,3 for the lag calculations. We
estimated the uncertainties in the lag measurements through
simulations as follows: For each of these bursts, we generated
simulated light curves in CH\,1 and CH\,3 using a two sided Gaussian
function whose left-width is set by the rise time of the initial
spike, the right-width is set by the decay time, and the amplitude
is set to the maximum rate in the particular energy band. The peak
time of the simulated CH\,3 curve is shifted by the measured lag
value of the burst. Using the same cross calibration method, we
calculated the lag between the two channels. For each burst, we
iterated this procedure 100 times and calculated the
root-mean-square value of the difference between the lag calculated
with the simulated light curves with respect to the measured value,
which was used as the uncertainty in the time lag.

\begin{figure}
\centering
\includegraphics*[scale=0.4]{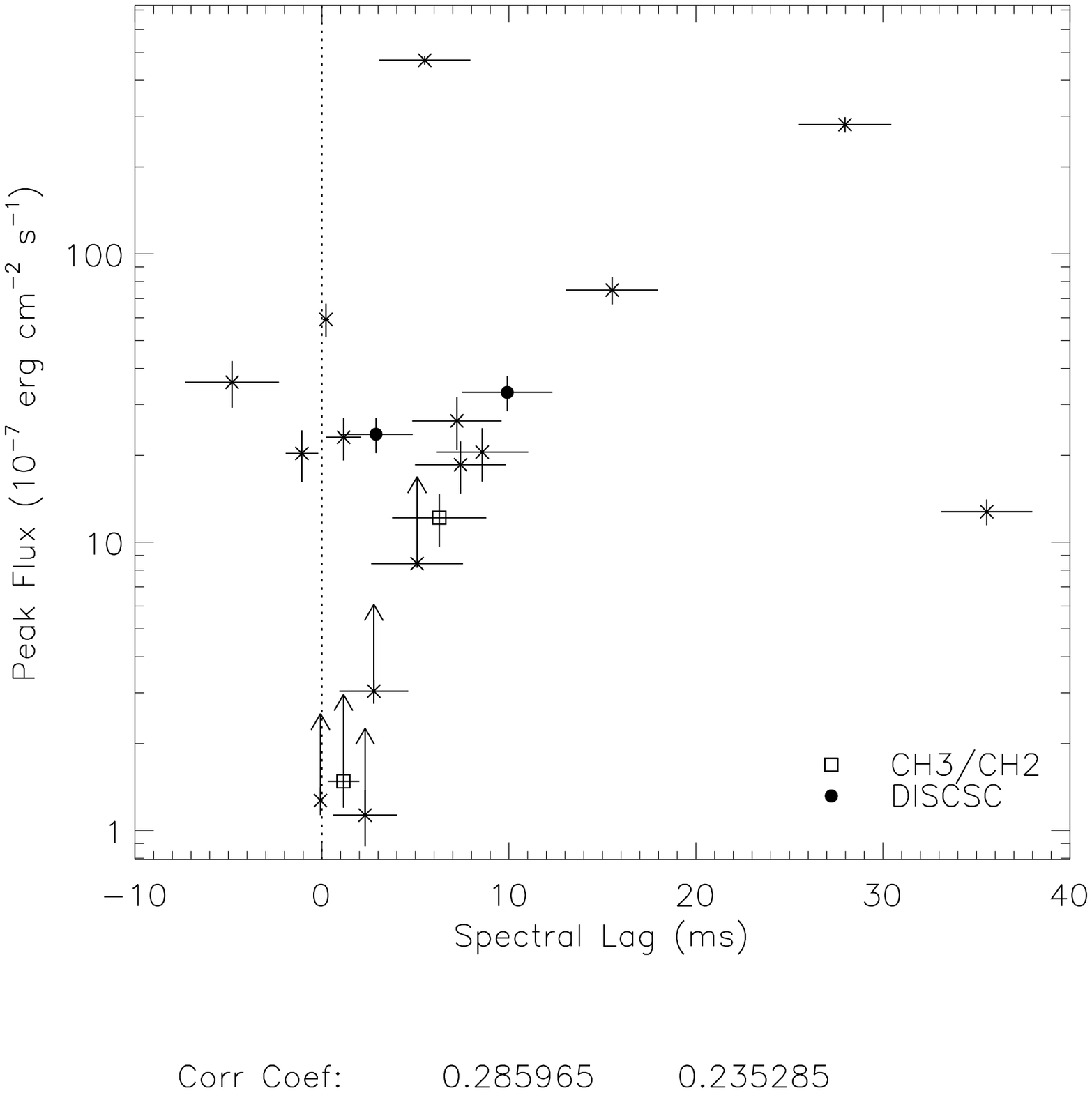}
\caption{Spectral lag vs.~peak flux.  The lags are between CH\,1 and
  CH\,3 of TTE data, and peak flux is in 30$-$1500\,keV (from
  Table~\ref{specanlysis}). The arrows show CONT events' peak flux,
  which should be taken as the lower limits. Squares show two events
  with lags calculated using CH\,2 and CH\,3.  Filled circles show
  events for which DISCSC data were used for the lags. No significant
  correlation was found (correlation coefficient, $R = 0.29$, chance
  probability $P = 0.24$.)} \label{lagfl}
\end{figure}

For most of the bursts the spectral lags of the spikes
were less than 10~ms (see Table~\ref{duration}). Although no
redshift values are known for any of these GRBs, the lag values are
consistent with those found for nearby short GRBs \citep{gehrels06},
if we assume typical redshift values for short GRBs of $z \lesssim
1$ \citep{berger11} for these events. We note that although our
sample size is small, there is no significant correlation between
the peak flux and the spectral lags for the spikes (see
Figure~\ref{lagfl}), consistent with the properties of short GRBs
\citep{norris00}.
The spectral lags for the spikes of most of these events are also
found in the BATSE spectral lag database \citep{hakkila07}.  There,
they used 64-ms DISCSC data to calculate the lags, while we used
8-ms TTE data to obtain our results.  The values are nonetheless
mostly consistent within uncertainties.

For the EE components, we were only able to calculate the lags for
four events with pronounced peaks in the EE light curves; those are
Trigger numbers 1997, 2436, 6782, and 7647. The lag values are
15.8$\pm$21.1, 48.1$\pm$21.0, 87.7$\pm$19.7, and
$-$22.0$\pm$21.2\,ms, respectively. Two of them show no lags, while
for the other two the lags are marginally larger than those of the
spikes. These lags are comparable to the average lag of long GRBs
\citep[$\sim$50\,ms;][]{norris00}.

\begin{table}
\centering \caption{Durations of the spikes and the extended
emission, $T_{90}$ from the BATSE GRB Catalog, and spectral lags of
the spikes for the 19 GRBs with EE. $T_{\rm Spike}$ and $T_{\rm EE}$
were estimated using 64-ms resolution data.}
\begin{tabular}{cccccc}
\hline
Trig$\#$& GRB name & $T_{\rm Spike}$ & $T_{\rm EE}$& $T_{90}$ & Spectral Lag\\
&& (s) & (s)  & (s) & (ms) \\
\hline
0575 & 910725 &  0.53  & 85.60 &0.41  &   7.41 $\pm$  2.43\\
 0603 & 910802 &  1.22  & 133.3 &1.47  &   2.89 $\pm$  1.96\\
 0906 & 911016 &  0.53  & 112.2 &0.40  &   7.22 $\pm$  2.39\\
 1088 & 911119 &  0.18  & 81.39 &0.19  &   9.91 $\pm$  2.41\\
 1719 & 920722 &  0.59  & 12.14 &1.05  &   8.57 $\pm$  2.46\\
 1997 & 921022 &  1.04  & 116.7 &60.22 &   5.09 $\pm$  2.45 \\
 2436 & 930709 &  4.86  & 29.20 &33.28 &   0.22 $\pm$  0.16 \\
 2611 & 931931 &  3.71  & 12.01 &12.21 &  27.97 $\pm$  2.47 \\
 3611 & 950531 &  2.75  & 55.29 &3.52  &  $-$0.08 $\pm$  0.06\\
 3940 & 951211 &  0.67  & 38.46 &0.58  &  $-$1.07 $\pm$  0.87\\
 5592 & 960906 &  0.45  & 87.04 &0.48  &   1.16 $\pm$  0.94\\
 5634 & 961017 &  1.42  & 45.33 &1.15  &  35.55 $\pm$  2.43\\
 6385 & 970918 &  0.29  & 14.66 &0.90  &   6.27 $\pm$  2.51\\
 6569 & 980112 &  1.22  & 81.89 &0.94  &   2.31 $\pm$  1.69\\
 6782 & 980525 &  1.37  & 22.53 &39.68 &   1.15 $\pm$  0.84 \\
 7063 & 980904 &  0.45  & 110.6 &0.13  &   2.78 $\pm$  1.84\\
 7446 & 990304 &  3.45  & 16.27 &13.57 &  15.52 $\pm$  2.45 \\
 7599 & 990605 &  1.00  & 24.19 &6.34  &  $-$4.80 $\pm$  2.50\\
 7647 & 990712 &  0.75  & 78.85 &67.8  &   5.50 $\pm$  2.44\\\hline
\end{tabular}
\label{duration}
\end{table}

\subsection{Spectral Analysis}

We also analyzed the spectra of the initial spikes and the EE
components of these GRBs. We used MER data (or CONT data in case
there is no MER data). Since the lowest one or two channels of the
MER/CONT data are usually below the electronic threshold, and the
highest channel is an energy overflow channel, we used a total of
about 14 energy channels ($\sim$30$-$1800\,keV) for each spectral
analysis.

We performed the spectral fitting with the RMFIT\footnote[1]{R. S.
Mallozzi, R. D. Preece, \& M. S. Briggs,
  ''RMFIT, A Lightcurve and Spectral Analysis Tool,'' \copyright~
  Robert D. Preece, University of Alabama in Huntsville.} version
  4.0rc1. For each event, we analyzed the spectra of the initial
spike and of the EE separately.  The durations of the EE used for
the analysis were matched to the time intervals where the excess
emission was identified by the search algorithm. We employed three
photon models: a single power-law model (PWRL), Comptonized (power
law with exponential cutoff) model with $E_{\rm peak}$
parametrization\footnote[2]{$f(E) = A \exp[-E(2+\lambda)/E_{\rm
peak}](E/E_{\rm piv})^\lambda$} (COMP), and the empirical GRB
function \citep[BAND;][]{band93}. We determined the simplest,
statistically well-fit model for the spectra of the spike and the
EE. We also took into account the spectral parameter constraints. We
present the spectral analysis results with the best model for each
component in Table~\ref{specanlysis}.

We find that either PWRL or COMP was sufficient to describe all of the
spikes as well as the EE components, with most of the spikes (EE) are
best fit with the COMP (PWRL) model. Based on the hardness ratios
(calculated with the photon flux of the best fit model, see
Table~\ref{specanlysis}), the spectra of the EE components are usually
softer, in accord with previous works in literature. In only one
  case (Trigger number 5634), the EE component is harder than the
  spike. We note that the resulting $\chi^2$ of the spike spectrum fit
  of this event is rather large (22.4 for 11 degrees of freedom,
  dof). This is due to some low-energy feature in the data, which
  likely comes from the spectral evolution within the spike. Although
  we do not present time-resolved spectral analysis here, some of
  these 19 events show significant spectral evolution in both the
  spikes and EE \citep[see also][]{kaneko06}, which result in such
  large $\chi^2$ values in time-integrated spectral modeling. In
  particular, the spectral data of trigger number 5634 could
  statistically be better described with models containing
  multi-spectral components, such as blackbody plus COMP. This,
  however, does not affect the total fluence values significantly and
  hence does not change the hardness ratio estimate.

We find a significant anti-correlation between the duration and the
hardness ratio of both spikes and EE combined (correlation
coefficient, $R = -0.63$, and chance probability, $P = 2.55\times
10^{-5}$). The spike spectra are on the hard side in terms of $E_{\rm
peak}$ values compared to the mean value of $\sim$320\,keV derived
from time-integrated spectra of bright BATSE GRBs
\citep{kaneko06}. The mean value of the PWRL indices for EE is
$-2.21\pm0.30$. The fact that larger fraction of the EE components are
better fit with PWRL could be due to either the real intrinsic soft
nature or the low statistics. The energy fluence ratios of the
EE to spike, in the 30$-$1500\,keV range vary from $\sim$0.1 to
$\sim$40 but most of them have values around one to a few (see
Table~\ref{specanlysis}).

We note that in some cases, EE spectra did not provide
well-constrained parameters or could not be fitted at all due to low
statistics. The EE of these GRBs were identified only in one energy
channel (mostly CH\,1 but one GRB in CH\,2), and even though the SNR
in the identified energy channel is sufficiently high, their
broadband spectra did not afford adequate total SNR for the
parameter constraints. Nevertheless, these spectra provide reliable
estimates on their relative hardness and energy fluence. It should
also be noted that the EE spectrum of another event (Trigger 603) is
fit with COMP with $E_{\rm peak} = 1126$\,keV and $\lambda < -2$,
indicating that the model describes a power law with exponential
{\it increase} above the $E_{\rm peak}$.  This is due to excess
emission at higher energies.

\begin{table*}
\center \scriptsize \setlength{\tabcolsep}{0.025in} \caption{Summary
of spectral fit results of 19 GRBs with EE.}
\begin{tabular}{ccccccccccccc}
\hline
Trig$\#$&GRB &Data &Component&Time interval&Model&A$\rm{^{a}}$&$E\rm{_{peak}}$&$\lambda$ &$\chi^{2}$/dof&Energy$\rm{^{b}}$&F$\rm{_{peak}^{c}}$&$HR_{3/2} \rm{^{d}}$\\
&Date&Type& &(s) & & &(keV) &  & &Fluence & & \\
\hline
0575&910725&MER &Spike &0.02 : 0.55&COMP&10.0 $\pm$ 1.2&413 $\pm$ 81 &$-$0.82 $\pm$ 0.17&19.5/11 & 1.76 $\pm$ 0.07&18.6 $\pm$ 3.8& 1.50 $\pm$ 0.12 \\
       &&&EE&4.41 : 90.01 &PWRL  &0.20 $\pm$ 0.04 & - &$-$2.57 $\pm$ 0.26&15.2/12 &11.25 $\pm$ 1.37&0.60 $\pm$ 0.22& 0.46 $\pm$ 0.15 \\
0603&910802&MER&Spike&0.03 : 1.25&COMP&7.3 $\pm$ 1.1&340 $\pm$ 99&$-$1.26 $\pm$ 0.18&11.1/11&3.07 $\pm$ 0.11&23.7 $\pm$ 3.3& 1.02 $\pm$ 0.07 \\
      &&&EE&5.68: 139.04&COMP&0.27 $\pm$ 0.03&1126 $\pm$ 240&$-$2.57 $\pm$ 0.18&11.4/11&25.42 $\pm$ 2.53&6.8 $\pm$ 2.2& 0.45 $\pm$ 0.09 \\
0906&911016&MER&Spike&0.02: 0.55&COMP&12.7 $\pm$ 1.4&420 $\pm$ 55&$-$0.37 $\pm$ 0.18&14.2/11&2.20 $\pm$ 0.07&26.3 $\pm$ 5.5& 1.86 $\pm$ 0.16 \\
      &&&EE&5.10: 117.27&PWRL&0.16 $\pm$ 0.04 &-&$-$1.42 $\pm$ 0.23&8.2/12&8.45 $\pm$ 1.91 &5.9 $\pm$ 2.2& 1.11 $\pm$ 0.40 \\
1088&91119&MER&Spike&0.03: 0.20&COMP&17.3 $\pm$ 4.3&282 $\pm$ 48&$-$0.37 $\pm$ 0.34&9.9/11&0.72 $\pm$ 0.04&33.1 $\pm$ 4.6& 1.61 $\pm$ 0.19 \\
      &&&EE&10.28: 91.67&PWRL&0.25 $\pm$ 0.03&-&$-$2.06 $\pm$ 0.19&5.7/12&10.07 $\pm$ 1.30&6.6 $\pm$ 1.9& 0.69 $\pm$ 0.14\\
1719&920722&MER&Spike&0.02: 0.62&COMP&8.5 $\pm$ 0.7 &548 $\pm$ 63&$-$0.14 $\pm$ 0.18&14.4/9&2.05 $\pm$ 0.07&20.5 $\pm$ 4.3& 2.55 $\pm$ 0.26 \\
              && &EE$^{\ast}$  &46.66 : 58.80& - & - & - & - & - & - & - & - \\
1997&921022&CONT  &Spike  &$-$2.75 : 3.39&PWRL  &3.9 $\pm$ 0.1 & -&$-$1.76 $\pm$ 0.04 &10.2/11& 10.93 $\pm$ 0.27 &102 $\pm$ 5& 0.85 $\pm$ 0.03 \\
       &&  &EE &3.39 : 120.13 &PWRL    &3.51 $\pm$ 0.04 & - &$-$2.09 $\pm$ 0.02&13.2/11  &204.40 $\pm$ 2.22 &20.4 $\pm$ 2.6& 0.64 $\pm$ 0.01 \\
2436&930709&MER&Spike&0.03: 4.90&COMP&15.9 $\pm$ 0.5&418 $\pm$ 16&$-$0.54 $\pm$ 0.05&20.6/11&25.24 $\pm$ 0.24&59.2 $\pm$ 7.9& 1.58 $\pm$ 0.03 \\
      &&&EE&8.85: 38.05&COMP&70.8 $\pm$ 0.2&354 $\pm$ 20&$-$1.20 $\pm$ 0.04&15.5/11&70.59 $\pm$ 0.56&17.0 $\pm$ 4.6& 1.01 $\pm$ 0.02 \\
2611&931031&MER&Spike&0.03: 3.74&COMP&42.9 $\pm$ 0.5&658 $\pm$ 24&$-$1.18 $\pm$ 0.02&65.2/10&62.80 $\pm$ 0.33&280 $\pm$ 17& 1.30 $\pm$ 0.01 \\
      &&&EE&5.82: 17.83&PWRL&1.58 $\pm$ 0.06&-&$-$2.16 $\pm$ 0.06&16.4/11&9.76 $\pm$ 0.31&6.8 $\pm$ 2.1& 0.65 $\pm$ 0.04 \\
3611&950531&CONT&Spike&$-$3.07: 7.17&PWRL&0.4 $\pm$ 0.1&-&$-$3.20 $\pm$ 0.26&14.1/12&5.70 $\pm$ 0.79&1.27 $\pm$ 0.14& 0.23 $\pm$ 0.08 \\
      &&&EE&19.46: 74.75&PWRL&0.2 $\pm$ 0.1&-&$-$2.77 $\pm$ 0.65&4.4/12&7.01 $\pm$ 2.38&0.34 $\pm$ 0.12& 0.34 $\pm$ 0.27 \\
3940&951211&MER  &Spike&0.03 : 0.70&COMP&16.5 $\pm$ 2.6&231 $\pm$ 24 &$-$0.73 $\pm$ 0.18 &15.9/11&2.50 $\pm$ 0.07&20.3 $\pm$ 4.1& 1.22 $\pm$ 0.07 \\
      &&  &EE &4.89 : 43.35&PWRL    &0.19 $\pm$ 0.04 & -&$-$2.65 $\pm$ 0.28&15.9/12 &5.07 $\pm$ 0.73&2.3 $\pm$ 1.6& 0.43 $\pm$ 0.14 \\
5592&960906&MER&Spike&0.02: 0.47&PWRL&3.8 $\pm$ 0.3&-&$-$1.48 $\pm$ 0.11&7.0/12&0.76 $\pm$ 0.07&23.1 $\pm$ 3.9& 1.01 $\pm$ 0.14 \\
     && &EE&10.01: 97.05&PWRL&0.19 $\pm$ 0.04 &-&$-$1.64 $\pm$ 0.27&17.2/11&7.58 $\pm$ 1.39&3.8 $\pm$ 2.0& 0.88 $\pm$ 0.27 \\
5634&961017&MER&Spike&0.03: 1.45&COMP&57 $\pm$ 12&57 $\pm$ 3&$-$1.10 $\pm$ 0.15&22.4/11&6.87 $\pm$ 0.21&12.8 $\pm$ 1.3& 0.40 $\pm$ 0.02 \\
      && &EE &5.04 : 50.33 &PWRL& 0.07 $\pm$ 0.03& - &$-$1.19 $\pm$ 0.31&11.3/10&1.62 $\pm$ 0.63&10.0 $\pm$ 2.8& 1.42 $\pm$ 0.86 \\
6385&970918&MER &Spike &0.02 : 0.31&COMP  &29.3 $\pm$ 7.1&265 $\pm$ 24&0.81 $\pm$ 0.38&15.5/10 &1.41 $\pm$ 0.06&12.2 $\pm$ 2.5& 2.46 $\pm$ 0.28 \\
      &&  &EE&22.34 : 37.02 &PWRL &0.13 $\pm$ 0.05 & - &$-$1.99 $\pm$ 0.68&14.7/11&0.92 $\pm$ 0.37&5.3 $\pm$ 2.6& 0.66 $\pm$ 0.46 \\
6569&980112 &CONT &Spike  &$-$1.66 : 4.48&PWRL &0.7 $\pm$ 0.1 &- &$-$1.81 $\pm$ 0.22&13.0/12&1.94 $\pm$ 0.27&1.13 $\pm$ 0.25& 0.81 $\pm$ 0.19 \\
      && &EE&18.82 : 100.74 &PWRL &0.33 $\pm$ 0.07 & - &$-$1.99 $\pm$ 0.34 &15.4/12  &12.84 $\pm$ 2.70 &0.41 $\pm$ 0.15& 0.70 $\pm$ 0.26 \\
6782&980525 &CONT &Spike  &$-$3.58 : 2.56&PWRL &0.4 $\pm$ 0.1 &  -&$-$1.34 $\pm$ 0.25&8.2/12 & 1.12 $\pm$ 0.25 &1.48 $\pm$ 0.28& 1.26 $\pm$ 0.45 \\
      && &EE&20.99 : 43.52 &PWRL &0.50 $\pm$ 0.05 & -&$-$1.68 $\pm$ 0.16&15.0/12& 5.08 $\pm$ 0.54 &0.99 $\pm$ 0.25& 0.94 $\pm$ 0.16 \\
7063&980904&CONT &Spike  &$-$2.88 : 3.26 &PWRL &0.7 $\pm$ 0.1 & -&$-$1.19 $\pm$ 0.12&14.4/12  &2.09 $\pm$ 0.27&3.04 $\pm$ 0.29& 1.35 $\pm$ 0.29 \\
      && &EE&21.70 : 132.29 &PWRL &0.13 $\pm$ 0.07 & - &$-$2.59 $\pm$ 0.67&6.8/12 &8.96 $\pm$ 2.76&0.78 $\pm$ 0.27& 0.41 $\pm$ 0.34 \\
7446&990304&MER&Spike&0.02: 3.48&COMP&30.8 $\pm$ 1.1&224 $\pm$ 12&$-$1.44 $\pm$ 0.03&14.6/10&34.49 $\pm$ 0.24&74.8 $\pm$ 8.1& 0.82 $\pm$ 0.01 \\
        &&&EE&5.99: 22.26&PWRL&0.71 $\pm$ 0.05&-&$-$2.29 $\pm$ 0.11&10.3/11&6.11 $\pm$ 0.36&2.5 $\pm$ 1.6& 0.51 $\pm$ 0.06 \\
7599&990605 &MER &Spike &0.03 : 1.04&COMP &5.3 $\pm$ 0.5&593 $\pm$ 91&$-$0.15 $\pm$ 0.22&4.5/10 &2.25 $\pm$ 0.10&35.8 $\pm$ 6.6& 2.51 $\pm$ 0.31 \\
      && &EE&14.30 : 38.49&PWRL&0.02 $\pm$ 0.04& - &$-$4.33 $\pm ^{1.23}_{1.72}$ &13.7/11&2.14 $\pm$ 2.37&7.9 $\pm$ 2.3& 0.12 $\pm$ 0.32 \\
7647&990712&MER&Spike&0.02: 0.78&COMP&73.1 $\pm$ 0.7&1230 $\pm$ 29&$-$0.40 $\pm$ 0.02&25.7/10&29.02 $\pm$ 0.19&469 $\pm$ 16& 2.25 $\pm$ 0.03 \\
      &&&EE&4.50: 83.35&PWRL&1.4 $\pm$ 0.1& - &$-$1.96 $\pm$ 0.13&8.6/10&52.13 $\pm$ 4.49&8.1 $\pm$ 2.0& 0.66 $\pm$ 0.09 \\
\hline
\end{tabular}
 \footnotesize{ \begin{flushleft}  All uncertainties are 1$\sigma$.\\
$\rm{^{a}}$ in units of $\rm{10^{-3}~phontons~cm^{-2}~s^{-2}~keV^{-1}}$ \\
$\rm{^{b}}$ in units of $\rm {10^{-7}~erg~cm^{-2}}$ and calculated in the 15$-$350~keV range \\
$\rm{^{c}}$ in units of $\rm {10^{-7}~erg~cm^{-2}~s^{-1}}$ and calculated in the 30$-$1500~keV range with 64-ms (MER) or 2-s (CONT) resolution \\
$\rm{^{d}}$ calculated in the 50 $-$ 100~keV and 100 $-$ 300~keV ranges\\
$\rm{^{\ast}}$ too dim to perform the spectral fit.
\end{flushleft}}
\label{specanlysis}
\end{table*}

\section{Discussion}
Based on a systematic search of 296 GRBs from the BATSE archive, we
identified 19 GRBs with EE that comply with our criteria with
sufficient statistics. 11 of these events are classified as short
  and 8 were long GRBs, based on the traditional dividing line of
$T_{90} = 2$\,s. The fraction of the BATSE GRBs with EE is
  $\sim$7\% of the total number of bursts consisting of initial short
  spikes, and having valid background data.  It has previously been
  reported that the fraction of the {\it Swift} BAT short GRBs with
EE corresponds to 2\% of the second BAT catalog \citep{sakamoto11},
and 25\% of BAT short GRBs \citep{norris10}. We caution that
  these statistics are not directly comparable since the criteria and
  methods used for the searches as well as for the parent sample
  selection were different. Nevertheless, some discrepancy in
  the EE detection frequencies is expected due to the differences in
the instrumental responses and triggering algorithms of BATSE and
  BAT. BAT triggers on smaller fraction of short GRBs ($\sim$10\% of
all BAT GRBs) compared to the BATSE sample but more sensitive to the
softer EE components once they are triggered. Moreover, our statistics
regarding the fraction of short GRBs with EE should be taken only as a
lower limit.  This is because the detection of EE components depends
on the detection threshold of the observing instruments as pointed out
by \citet{norris10}, since most of the EE components are very
weak. Especially considering relatively lower sensitivity of BATSE LAD
in energies $\lesssim 100$\,keV and the soft nature of the EE, as well
as relatively noisy background in BATSE data, possibly a larger
fraction of the BATSE short GRBs are accompanied by EE. We note,
however, \citet{norris10} estimates that such EE would have been
detected in as much as 50\% of {\it Swift} BAT short GRBs if it had
been present, after taking into account the the detector sensitivity,
suggesting that there is a physical threshold for the EE and only
$\sim25\%$ of short GRBs are really associated with EE.

The spectral lags of the spikes of the 19 GRBs are small with most
of them $\lesssim$ 10\,ms, consistent with other short-duration
GRBs. For the EE, we were able to determine the lags for four
events, two of which indicate no spectral lags and the other two
comparable to those of long GRBs. In the case of GRB 060614, the
spectral lag for the EE was found to be insignificant similar to the
spike, which may be suggestive of the same physical origin for both
of the components. In contrast, our results may point to different
origins for at least some of the EE, possibly caused by milder
outflow with smaller Lorentz factor than the origin of the spikes.

The spike and EE components are both well described with either a
single power law model or a Comptonized model. In all cases except
one, the EE components are softer than the spike based on their
hardness ratios. The energy fluence ratio of the two components
(E$_{\rm EE}$/E$_{\rm spike}$) spans a wide range, varying from
$\sim$0.1 to $\sim$40, as was also found by \citet{norris06}, who also
searched visually GRBs with EE within the BATSE archive and identified
eight GRBs with EE. Among the eight events of \citet{norris06}, only
two were identified in our systematic search (Triggers 1997 and
7647). They reported that EE component is always softer than the
initial spike as determined by their hardness ratios.  Our results
confirm the soft nature of the EE in general, although several GRBs
in our sample have EE that are comparably as hard as, or harder than
the corresponding spikes.

Based on the spectral analysis, at least four events (Trigger
numbers 1088, 1997, 6385, and 6569) are spectrally very similar to
GRB\,060614.  The spectral lags of the spikes of all these events
are of the order of a millisecond, also similar to 060614 although
our uncertainties are smaller. All of them consist of a spike
$\lesssim$ 1.5\,s, and the EE of three of them are $\sim 100$\,s
while the other has a short EE of about 15\,s. The peak (isotropic)
luminosity of 060614 spike was $1.5\times10^{50}$\,erg\,s$^{-1}$. If
we assume the same cosmological redshift as 060614 (z = 0.125), the
peak luminosity estimates for these four events are in the range of
$4.5\times10^{48}-4\times10^{50}$\,erg\,s$^{-1}$.

The nature of the EE may be explained by the interaction of
relativistic outflow with the circumburst medium (i.e, early
afterglow) or late time central engine activity. However, the temporal
variations in the light curves of some bright EE increase the
possibility of ongoing central engine activity.

The collapsar model for long bursts could produce a burst consisting
of a short-hard spike and EE in various scenarios. \citet{zhang03}
showed that the short-hard ``precursor" is created due to the
relativistic jet breakout and subsequent interaction with the
surrounding stellar wind.  The following long, main burst component
is then created by the internal shocks.  The relative brightness of
the short precursor and the main burst component can vary greatly,
as it depends on many physical parameters at two different origins.
Another model was proposed by \citet{laz10}, in which the EE in
short GRBs could be created in collapsar viewed off axis. A critical
piece of observational evidence for these model is a supernova
association, which extensive follow-up optical observations have
excluded for some short GRBs with EE
\citep{hjorth05,galyam06,fynbo06,perley09}. These models also expect
that the bursts take place in active star-forming regions of the
host galaxies, which have also been ruled out for many short GRBs
whose host galaxies have been identified
\citep{gehrels05,villa05,berth05}. Unfortunately, there was no
follow-up observations in longer wavelengths for the GRBs with EE in
our sample to search for such associations.

Alternatively, \citet{mac05} have proposed that accretion-induced
collapse of a neutron star into a black hole, similar to Type Ia
supernova, may also be able to produce EE, if the jet from the
accretion interacts with the envelope of a companion giant star. In
this case, a supernova may not be created and the host galaxies can
be both early and late types due to the widely-varying timescale in
which the sequence of the events occurs.  The high variability
observed in some of the EE, however, may be difficult to reconcile
with this model.

On the other hand, short GRBs with EE could also be produced in the
context of the merger model of short GRBs, since the central engine
may produce a relativistic outflow made of two different components.
\citet{metzger08} and \citet{buc12} have shown that these GRBs could
be created by a protomagnetar (highly-magnetized, rapidly-rotating
neutron star), formed either in accretion-induced collapse of a
white dwarf, or in a merger of white dwarf binary or neutron star
binary. In this scenario, the initial spike of the GRB is produced
by the accretion powered jet, while the EE originates from a jet
powered by the spin down of the protomagnetar, as it breaks out of
the confined envelope of ejected mass. The large variation in the
observed energy ratio of the spike and EE may also be explained by a
possible wide range in initial angular momentum of the system.
Although no bright supernova would accompany such a collapse or
merger process, a smaller scale supernova would be expected around 1
day after the burst.  A peak in optical wavelength similar to this
signature has been observed in at least one short GRB with bright EE
(GRB\,080503), albeit its inconclusive nature \citep{perley09}.

Recently, \citet{barkov11} suggested a two component jet
model in the context of compact binary merger to explain short GRBs
with EE. In this model, the short spike is due to a jet powered by
neutrino heating ($\nu\tilde{\nu}$ annihilation)
 while the EE results from a jet produced by the
Blandford-Znajek (BZ) mechanism. The opening angle of the
neutrino-powered jet ($\theta_{\nu\tilde{\nu}}$) is typically of the
order of $\sim$ 0.1 \citep[for Lorentz factor, $\Gamma >
100$;][]{aloy05}, which can be much larger than the BZ-powered jet,
$\theta_{BZ} \sim 1/\Gamma$. Then, the wide range of the energetic
ratios of the short spikes and EE can be explained by various
off-axis viewing positions.

\citet{barkov11} also estimated the ratio of the opening
angles ($\theta_{BZ}/\theta_{\nu\tilde{\nu}}$) using the observed
fluence of the spike and EE in the case of GRB 060614, based on the
relation:
\begin{equation}
{{L_{BZ}}\over{L_{\nu\tilde{\nu}}}}~\left(\frac{\theta_{\nu\tilde{\nu}}}
{\theta_{BZ}} \right)^2 {{t_{BZ}}\over{t_{\nu\tilde{\nu}}}}=
{{E_{\rm EE}}\over{E_{\rm spike}}}
\end{equation}
Here, $E_{\rm EE}$ and $E_{\rm spike}$ are the observed energy
fluence, and $t_{\nu\tilde{\nu}}$ and $t_{BZ}$ are durations of the
spike and the EE, respectively. They estimate the luminosities of
the both jets to be $L_{BZ} \approx 10^{48}$\,erg\,s$^{-1}$ and
$L_{\nu\tilde{\nu}} \approx 3\times10^{50}$\,erg\,s$^{-1}$, assuming
typical physical parameters of the progenitor, such as the black
hole mass, spin parameter, accretion rate, and disc viscosity.  For
GRB 060614, $\theta_{BZ}$/$\theta_{\nu\tilde{\nu}} \sim 0.1.$ Using
this equation, we estimated the ratios of jet opening angles of our
18 GRBs with EE (see Table~\ref{modelresults}). Trigger number 1719
was excluded here because the energy fluence value for the EE was
not obtainable due to its low statistics.  The fluence values used
for these estimates were calculated in the energy range of
15$-$350\,keV, so as to allow direct comparison with the values
obtained for GRB\,060614.
\begin{table}
\centering \caption{Estimated ratios of opening angles of the two
jet components.}
\begin{tabular}{cccc}
\hline Trig$\#$& GRB name & $E_{\rm EE}/E_{\rm Spike}$ &
$\theta_{BZ}$/$\theta_{\nu\tilde{\nu}}$ \\
 & & (15--350keV) & \\
\hline
 0575 & 910725   &   6.38  &  0.29 \\
 0603 & 910802   &   8.29  &  0.21 \\
 0906 & 911016   &   3.84  &  0.43 \\
 1088 & 911119   & 13.98  &  0.33 \\
 1997 & 921022   &  18.70 &  0.06 \\
 2436 & 930709   &   2.80  &  0.08 \\
 2611 & 931931   &   0.16  &  0.26 \\
 3611 & 950531   &   1.23  &  0.12 \\
 3940 & 951211   &   2.03  &  0.31 \\
 5592 & 960906   &   9.92  &  0.26 \\
 5634 & 961017   &   0.24  &  0.67 \\
 6385 & 970918   &   0.66  &  0.51 \\
 6569 & 980112   &   6.64  &  0.08 \\
 6782 & 980525   &   4.53  &  0.05 \\
 7063 & 980904   &   4.29  &  0.12 \\
 7446 & 990304   &   0.18  &  0.30 \\
 7599 & 990605   &   0.95  &  0.29 \\
 7647 & 990712   &   1.95  &  0.44 \\
\hline
\end{tabular}
\label{modelresults}
\end{table}
We find that the angle ratios,
$\theta_{BZ}$/$\theta_{\nu\tilde{\nu}}$, span a wide range from 0.05
to 0.67. The ratio is always less than one, meaning that the BZ jet
(responsible for EE) is more narrowly collimated than the neutrino
jet (spike). This also implies that (from the above equation) the
ratio of average flux values are always less than the ratio of the
luminosity estimates of the two components; namely, ${\bar{F}_{\rm
EE}}/{\bar{F}_{\rm spike}} < {L_{BZ}}/{L_{\nu\tilde{\nu}}}$. In
turn, this indicates either $L_{BZ} < 10^{48}$\,erg\,s$^{-1}$ or
$L_{\nu\tilde{\nu}} > 3\times10^{50}$\,erg\,s$^{-1}$.

Interestingly, both GRBs for which significant spectral lags in the
EE were found, have rather small angle ratios,
$\theta_{BZ}$/$\theta_{\nu\tilde{\nu}} < 0.1$. This may indicate
much wider neutrino-powered jets and hence the smaller $\Gamma$ for
these events. Moreover, although we see no significant correlation
between the angle ratios and the spectral lags of the spikes, the
event with the largest spike lag (Trigger 5634; 35.55$\pm$2.43\,ms)
has the largest angle ratio estimate. These are puzzling, however,
since larger lags are expected from less collimated jet structures,
if the spectral lags are solely due to the curvature effect. By
comparison, the angle ratio estimated for GRB 060614 was also
$\sim$0.1 but no spectral lag for neither the spike nor the EE was
found.  Also, one of our sample with no spectral lag in the EE
(Trigger 1997) has a small angle ratio of 0.06. In these cases, the
collimation angle for the neutrino jet may still need to be
sufficiently small, which in turn indicates much higher $\Gamma$ for
the BZ jet.

Finally, we note that some short GRBs with EE are observed to have
X-ray flare-like re-brightening in their afterglow light curves
\citep{berth05,mar11}. The X-ray flares are observed in lower energy
($\lesssim 10$\,keV), usually at around a few hundreds to
thousands of seconds after the gamma-ray prompt emission. They can
be very bright with relatively sharp rise and fall, and are normally
superimposed on a smoothly decaying light curve
\citep{chincarini10}.  Their temporal and spectral properties as
well as energetics are very similar to GRB prompt emission (at least
for long GRBs), suggesting an internal origin related to the central
engine activity. If the EE identified in our search extends down to
a few keV, they could resemble X-ray flares occurring at a very
early stage of the afterglow \citep[e.g.,][]{laparola06}.

\section*{Acknowledgments} 
We thank the referee for valuable comments and suggestions. This
project is founded by the Scientific and Technological Council of
Turkey (T\"UB\.ITAK grant number 109T755).


\begin{thebibliography}{}

\bibitem[Aloy, Janka \& M\"uller(2005)]{aloy05} Aloy, M. A., Janka, H.-T. \& M\"uller,
E., 2003, A\&A, 436, 273

\bibitem[Band et al.(1993)]{band93} Band, D. L., et~al., 1993, ApJ, 413, 281

\bibitem[Barkov \& Pozanenko(2011)]{barkov11} Barkov, M. V. \& Pozanenko, A. S., 
2011, MNRAS, 417, 2161

\bibitem[Barthelmy et al.(2005)]{berth05} Berthelmy, S. D., et al.,
2005, Nature 438, 994

\bibitem[Berger(2011)]{berger11} Berger, E., 2011, New Astronomy Reviews, 55, 1

\bibitem[Bucciantini et al.(2012)]{buc12} Bucciantini, N., et al.,
2012, MNRAS, 419, 1537

\bibitem[Chincarini et al.(2010)]{chincarini10} Chincarini, G., et
al., 2010, MNRAS, 406, 2113

\bibitem[Connaughton(2002)]{connaughton02} Connaughton, V.,  2002 ApJ 567, 1028.

\bibitem[Cucchiara et al.(2011)]{cucc11}Cucchiara, A., et al., 2011, ApJ, 736, 7

\bibitem[Della Valle et al.(2006)]{dv06} Della Valle, M., et al., 2006,
Nature, 444, 1050

\bibitem[Eichler et al.(1989)]{eichler89}Eichler, D., Livio, M., Piran, T., Schramm, D. N., 1989, Nature, 340, 126

\bibitem[Fishman et al.(1989)]{fis89}Fishman, G.J., et~al. 1989, in Proc. Gamma Ray Observatory Science Workshop,
ed. W. N. Johnson (Greenbelt: NASA/GSFC ), 39

\bibitem[Frederiks et~al(2004)]{frederiks04}Frederiks, D. D, et al., 2004, ASPC, 312, 197

\bibitem[Fruchter et al.(2006)]{fruchter06} Fruchter, A., et al., 2006, Nature, 441, 463.

\bibitem[Fynbo et al.(2006)]{fynbo06} Fynbo, J. P. U., et al., 2006,
Nature, 444, 1047

\bibitem[Gal-Yam et al.(2006)]{galyam06} Gal-Yam, A., et al., 2006,
Nature, 444, 1053

\bibitem[Gehrels et al.(2005)]{gehrels05} Gehrels, N., et al., 2005,
Nature, 437, 851

\bibitem[Gehrels et al.(2006)]{gehrels06} Gehrels, N., et al., 2006,
Nature, 444, 1044

\bibitem[Hakkila et al.(2007)]{hakkila07}Hakkila, J., et~al.,
2007, ApJS, 169, 62

\bibitem[Hjorth et al.(2005)]{hjorth05} Hjorth, J., et al., 2005,
Nature, 437, 859

\bibitem[Jakobsson et al.(2012)]{jakcobsson12}Jakobsson, P., et~al.,
2012, ApJ, 752, 62

\bibitem[Kaneko et al.(2006)]{kaneko06}Kaneko, Y., et~al., 2006, ApJ, 166, 298

\bibitem[Kouveliotou et al.(1993)]{kouveliotou93} Kouveliotou, C., et al., 1993, ApJ, 413, 101

\bibitem[La Parola et al.(2006)]{laparola06} La Parola, V., et
al., 2006, A\&A, 454, 753

\bibitem[Lazzati et al.(2001)]{laz01} Lazzati, D., Ramirez-Ruiz, E., Ghisellini, G. 
2001, A\&A, 379L, 39

\bibitem[Lazzati, Morsony \& Begelman(2010)]{laz10} Lazzati, D.,
  Morsony, B.J., \& Begelman, M.C. 2010, ApJ, 717, 239

\bibitem[Lazzati, Ramirez-Ruiz, \& Ghisellini(2001)]{laz01}Lazzati, D., Ramirez-Ruiz, E., \&
Ghisellini, G., 2001, A\&A, 379, L39

\bibitem[Levesque et al.(2010)]{levesque10} Levesque, E. M., Kewley,
  L. J., Berger, E., Zahid, H. J., 2010, AJ, 140, 1557

\bibitem[MacFadyen, Ramirez-Ruiz, \& Zhang(2005)]{mac05}MacFadyen,
  A. I., Ramirez-Ruiz, E., \& Zhang, W. 2005, arXiv:astro-ph/0510192v1

\bibitem[Mazets et~al.(2002)]{mazets02}Mazets, E. P. et al., 
2002, astro.ph, 9219

\bibitem[Margutti et al.(2011)]{mar11} Margutti, R., et al. 2011, MNRAS,
417, 2144

\bibitem[Metzger, Quataert, \& Thompson(2008)]{metzger08}Metzger, B. D., Quataert,
E., \& Thompson, T. A. 2008, 385, 1455

\bibitem[Montanari et al.(2005)]{montanari05} Montanari, E., Frontera,
  F., Guidorzi, C., Rapisarda, M., 2005, ApJ, 625L, 17

\bibitem[Narayan et~al.(1992)]{narayan92}Narayan, R., Paczynski, B., Piran, T., 1992, ApJ, 395L, 83

\bibitem[Norris \& Bonnell(2006)]{norris06} Norris, J. P. \& Bonnell, J. T., 2006, ApJ, 643, 266

\bibitem[Norris, Gehrels, \& Scargle(2010)]{norris10} Norris, J.P., Gehrels, N., \& Scargle, J.D.,
2010, ApJ, 717, 411

\bibitem[Norris, Marani, \& Bonnell(2000)]{norris00} Norris, J. P.,
  Marani, G. F., \& Bonnell, J. T., 2000, ApJ, 534, 248

\bibitem[Paczynski(1986)]{paczynski86} Paczynski, B., 1986, ApJ, 308L, 43

\bibitem[Perley et al.(2009)]{perley09} Perley, D. A., et al., 2005,
ApJ 696, 1871

\bibitem[Preece et~al.(2000)]{pre00}Preece, R. D., et~al. 2000, ApJ, 126, 19

\bibitem[Sakamoto et~al(2011)]{sakamoto11}Sakamoto, T. et~al., 2011, ApJS, 195, 2 

\bibitem[Villasenor et al.(2005)]{villa05}Villasenor, J. S., et
al., 2005, Nature, 437, 855

\bibitem[Woosley(1993)]{woosley93}Woosley, S. E., 1993, ApJ, 405, 273.

\bibitem[Zhang, Woosley, \& MacFadyen(2003)]{zhang03}Zhang, W., Woosley, S. E., \& MacFadyen, A. I., 2003, ApJ, 586, 356.




\end{thebibliography}
\end{document}